\documentstyle[preprint,revtex]{aps}
\begin{document}
\begin{title}
\begin{center}
Binding energies and nonradiative decay rates of $H\! ed\mu
$-molecular ions
\end{center}
\end{title}
\author{V.B.Belyaev,
O.I.Kartavtsev\dag,
V.I.Kochkin\ddag, E.A.Kolganova\ddag}
\begin{instit}
Bogolubov Laboratory of Theoretical Physics \\
Joint Institute for Nuclear Research\\
141980, Dubna, Russia\\
Fax: 7 096 21 65084\\
\dag{To whom correspondence should be addressed\\
e-mail:~oik@thsun1.jinr.dubna.su\\}
\ddag{Laboratory of Computing Techniques and Automation}
\end{instit}

\begin{abstract}
The method of hyperspherical "surface" functions has been applied
to the calculation of eigenvalues and eigenfunctions of
muonic molecular ions $^{3,4}\! H\! ed\mu $. Binding energies and
nonradiative decay rates for the states of the total angular momentum
$L=0,~1,~2$ have been obtained.
\end{abstract}

PACS number: 36.10.Dr

\vfil
\section{Introduction}

The reasons for investigation of
the charge-nonsymmetric muonic molecules like $H\! eH\mu $ are as
follows. First, a direct charge-exchange reaction from the
ground-state muonic hydrogen atom to helium nuclei is suppressed, the
transfer proceeds through the formation of the molecule in the intermediate
state. Hence, the kinetics of muons in media is defined to a large extent
by the probability of this process. Indeed, the role of the formation of
a muonic molecule in a charge-exchange reaction was confirmed in a number
 of experiments ~\cite{jones},~\cite{byst}, ~\cite{balin}, ~\cite{jacot}.

The measurement of the yield of $\gamma $-rays due to the decay of the
$H\! ed\mu $ molecules ~\cite{mats}, ~\cite{ishid} revives
interest in the investigation of this system. The experiment gives
evidence of an additional nonradiative decay channel. This possibility
was discussed in papers ~\cite{kino}, ~\cite{ger}, ~\cite{krav2}.

Next, eigenenergies of both usual molecules of media and muonic
molecules, as it follows from the results of ~\cite{krav},
{}~\cite{hara}, are comparable. For this reason one can expect an
active interaction of muonic molecules with media.

 As soon as the
charge-nonsymmetric molecules are produced, the possibility of nuclear
transitions arises. The investigation of a nuclear reaction at
typical mesomolecular energies has a fundamental importance due to
absence of any experimental data on the strong interaction of
charged particles in this energy range.

Qualitatively properties of the $H\! eH\mu $ system are defined as
follows. Coulomb interaction is not able to
bind the systems under consideration due to the repulsion in the $H\!
e\mu +H$ channel. Only a $3-$body resonant state can be formed. States
like that are supported by the attractive polarization potential in the
$H\mu +H\!  e$ channel and therefore are clustered.

The goal of this paper is to perform
systematical calculations of energy levels and nonradiative decay rates of
the $^{3,4}H\! ed\mu $ systems for all possible values of the total angular
momentum. Only the transition to the channel with the $H\! e\mu $ atom
in the ground state has been considered. Transitions from the molecular
states with $L=1,2$ to the channels with the $H\! e\mu $ atom in the $2s,
2p$ states will be suppressed due to exponentially small overlapping of
the initial and final state wave-functions.

The treatment of this problem met essential difficulties due
to necessity to describe the Coulomb three-body system above the
two-body threshold.
Some approaches to describing
these systems have been applied in ~\cite{hara},
{}~\cite{kino},
{}~\cite{ger}, ~\cite{krav2}.

The approach, using the
hyperspherical "surface" functions method ~\cite{mac},
{}~\cite{lin} has been
applied in this paper. The following advantages of this method
in treating the posed problems can be mentioned.
The method operates with a discrete set of coupled one-dimensional
differential equations. Physical boundary conditions for their
solution can be easily formulated. Moreover, coupling of channels
turns out to be rather small in our case and allows one to use the
decoupled one-level approximation. It is worthwhile to mention the
analogous calculation of the $LiH\mu $ and $BeH\mu $
molecules ~\cite{bel}.

The article is organized in the following way. The description
of the method will be given in the next section, section 3 contains
numerical results, section 4 - discussion and conclusion.

\section{Method}

The Hamiltonian of three charged particles in the Jacobi variables is:
\begin{equation}
H=-\Delta _{\displaystyle{\bf x}_{i}}-\Delta _{\displaystyle{\bf y}_{i}}+
\sum\limits_{s=1}^{3}\frac{q_{s}}{x_{s}},
\label{eq:h1}
\end{equation}
where
\begin{equation}\left.
\begin{array}{l}
{\bf x}_{i}=\sqrt{\displaystyle\frac{m_{k}m_{j}}{m(m_{k}+m_{j})}}({\bf r}_{k}
-{\bf r}_{j}),\vspace{.5cm}\\
{\bf y}_{i}=\sqrt{\displaystyle\frac{m_{i}(m_{k}+m_{j})}{m(m_{i}+m_{k}+m_{j})}}
({\bf r}_{i}-\displaystyle\frac{m_{j}{\bf r}_{j}+m_{k}{\bf
r}_{k}}{m_{j}+m_{k}}),
\label{eq:h2}
\end{array}\right.
\end{equation}
\begin{equation}
q_{i}=2Z_{j}Z_{k}\sqrt{\frac{m_{k}m_{j}}{m(m_{k}+m_{j})}}.
\label{eq:h3}
\end{equation}
${\bf r}_{i}, m_{i}, Z_{i}$ - coordinate, mass and charge of the
$i$-th particle. $\hbar ^2/(me^2),\ me^4/(2\hbar ^2)$ have been used
as length and energy units. Here $m$ is
arbitrary mass and was taken equal to the muonic mass. For
definiteness muon, hydrogen nucleus and
nucleus of the charge Z have been enumerated as particles with number 1, 2
and 3.

Coordinates
 $\rho ,\alpha _{i}, \theta _{i}$have been introduced by the relations
\begin{equation}\left.
\begin{array}{l}
x_{i}=\rho cos\displaystyle\frac{\alpha _{i}}{2},\\
y_{i}=\rho sin\displaystyle\frac{\alpha _{i}}{2},\\
cos\theta _{i}=\displaystyle\frac{({\bf x}_{i}\cdot {\bf
y}_{i})}{x_{i}y_{i}},\\
0\ \leq \ \alpha _{i}, \theta _{i}\ \leq \pi .
\label{eq:h4}
\end{array}\right.
\end{equation}
Below the notation $\Omega $ will be used for an arbitrary pair
$\alpha_i, \theta_i$.

Since the systems have two heavy and one light particles, it is reasonable
to assume that the main part of the total angular momentum is carried by
the pair of heavy particles. This is reason why the following form of the
solution of the Schr\" odinger equation has been used:
\begin{equation}
\Psi _{LM}({\bf x},{\bf y})=Y_{LM}({\bf\hat x}_1)\Phi _L(\rho ,\Omega ).
\label{eq:mom}
\end{equation}

Under these assumptions the Schr\" odinger equation for the
states with total angular momentum $L$ takes the form:
\begin{equation}
[-\frac{1}{\rho ^5}\frac{\partial }{\partial \rho} (\rho ^5\frac{\partial
}{\partial \rho} )- \frac{4}{\rho^2}\Delta_{\Omega
}+ \sum\limits_{s=1}^{3}\frac{q_{s}}{x_{s}}-E]\Phi _L(\rho ,\Omega )=0,
\label{eq:e5}
\end{equation}
where
\begin{equation}
\Delta_{\Omega }=\frac{1}{sin^2\alpha _{i}}
[\frac{\partial }{\partial \alpha_{i}} (sin^2\alpha _{i}\frac{\partial
}{\partial \alpha_{i}} )+
\frac{1}{sin\theta _{i}}\frac{\partial }{\partial \theta_{i}} (sin\theta
_{i}\frac{\partial }{\partial \theta_{i}})]
-\displaystyle\frac{L(L+1)}{4cos^{2}\alpha _1}.
\label{eq:e6}
\end{equation}
Following ~\cite{mac} and ~\cite{lin} "surface"
functions $\varphi _{n}(\Omega ;\rho )$
can be introduced as finite solutions of the
equation:
\begin{equation}
[\Delta_{\Omega }-\frac{\rho }{4}
\sum\limits_{s=1}^{3}\frac{q_{s}}{x_{s}}
+\lambda _{n}(\rho )]\varphi_{n}(\Omega ;\rho )=0.
\label{eq:e7}
\end{equation}
Expanding
the solution of the equation (\ref{eq:e5}) onto the set of the
hyperspherical "surface" functions:
\begin{equation}
\Phi _L(\rho ,\Omega )=
\rho ^{-5/2}\sum\limits_{n}u_{n}\varphi _{n}(\Omega ;\rho),
\label{eq:e8}
\end{equation}
one immediately comes to the system of one-dimensional equations
\begin{eqnarray}
[\frac{d^2}{d\rho ^2}-\frac{15}{4\rho ^2}-
\varepsilon _{n}(\rho)&+&E]u_{n}(\rho )+\nonumber\\
\sum\limits_{i}[Q_{ni}(\rho )\frac{d}{d\rho} &+&
\frac{d}{d\rho} Q_{ni}(\rho )-P_{ni}(\rho )]u_{n}(\rho )=0,
\label{eq:e9}
\end{eqnarray}
where
\begin{equation}
Q_{ni}(\rho )=\langle \varphi_{n}|\frac{\partial }{\partial \rho}
\varphi_{i}\rangle ,
\label{eq:e10}
\end{equation}
\begin{equation}
P_{ni}(\rho )=\langle \frac{\partial }{\partial \rho}
\varphi_{n}|\frac{\partial }{\partial \rho} \varphi_{i}\rangle ,
\label{eq:e11}
\end{equation}
\begin{equation}
\varepsilon _{n}(\rho)=\frac{4}{\rho^2}\lambda _{n}(\rho ).
\label{eq:e12}
\end{equation}
$\langle \ \cdot \ |\ \cdot \ \rangle$ means the integration on the
hypersphere over $d\Omega =sin^2\alpha _{i}d\alpha _{i}dcos\theta _{i}$.

One of the most complicated problems of this approach is the
computation of $Q_{ni}(\rho )$ and $P_{ni}(\rho )$, defined in
(\ref{eq:e10}) and (\ref{eq:e11}). By this reason, the following exact
expressions have been used:  \begin{equation} Q_{ni}=-\frac{1}{4}(\lambda
_{i}-\lambda _{n})^{-1}\langle
\varphi_{n}|\sum\limits_{s=1}^{3}\frac{q_{s}}{x_{s}} |\varphi_{i}\rangle,
\label{eq:q1}
\end{equation}
\begin{equation}
P_{ni}=-(Q^2)_{ni}.
\label{eq:p1}
\end{equation}
The form (\ref{eq:q1}), (\ref{eq:p1}) allow one to avoid the
calculation of the derivatives of the "surface" functions on the
parameter $\rho $ and use only already known matrix elements $V_{ni}(\rho )$
and eigenvalues $\lambda _{i}(\rho )$ of equation (\ref{eq:e7}).

The variational approach has been applied to solve equation
(\ref{eq:e7}). The "surface" functions have been chosen as a linear
combination of trial functions from the following set:
\begin{equation}\left.
\begin{array}{l}
\phi ^{(\sigma )}_{nl}(\alpha _{\sigma })
P_{l}(cos\theta _{\sigma }),\quad \sigma =2,3,\\
sin^l\alpha _{3}C^{l+1}_{n-l-1}P_{l}(cos\theta _{3}),\\
n>0,\ n>l\geq 0,
\end{array}\right.
\label{eq:s13}
\end{equation}
where
\begin{equation}\left.
\begin{array}{l}
\phi ^{(\sigma )}_{nl}(\alpha )=
R_{nl}(\displaystyle{|q_\sigma |\over n}
\rho cos{\displaystyle\alpha \over 2}),\\
R_{nl}(t)=exp(-t/2)t^{l}L_{n-l-1}^{2l+1}(t)
\end{array}\right.
\label{eq:s14}
\end{equation}
In equations (\ref{eq:s13}) and (\ref{eq:s14}) $P_{l}(x), L_{m}^{k}(x),
C_{n}^{m}(x)$ are the Legendre, Laguerre and Gegenbauer polynomials.
The set of trial functions has been chosen in the
form (\ref{eq:s13}) in order to describe properly the three-body
wave-function at both large and small interparticle distances.
The first line of (\ref{eq:s13}) will describe the system
separated into two clusters. In this case, one of the
clusters is a hydrogen-like atom and hydrogen-like functions
(\ref{eq:s14}) will be proper trial functions. The second line of
(\ref{eq:s13}) will describe the configuration with all three particles
close to each other.  In this case, the centrifugal term in (\ref{eq:e5})
dominates and eigenfunctions of the operator (\ref{eq:e6}) are used. The
set of trial functions (\ref{eq:s13})can be easily adjusted to the
different values of the parameter $\rho $. For this purpose numbers of
channel-type functions and hyperspherical harmonics have been changed with
changing $\rho $. It is necessary to emphasize that the dependence of the
numbers of the trial functions on the parameter $\rho $ has not been
exploited in analogous calculations. This dependence gives rise to more
flexibility of the basis and allows one to avoid numerical instabilities
when solving equation (\ref{eq:e7}).

As a result of the solution of equation (\ref{eq:e7}) eigenpotentials
$\varepsilon _{n}(\rho )$, $Q_{12}(\rho )$, $P_{12}(\rho )$ have been
obtained.
The properties of mesomolecules and transition rates
are mostly defined by the specific form of the
effective potentials $\varepsilon _{n}(\rho )$. The lowest
effective potential $\varepsilon _{1}(\rho )$ describes
asymptotically the decay channel $H+He\mu $ and is repulsive at all $\rho
$ values. The next effective potential $\varepsilon
_{2}(\rho )$ describes asymptotically the channel $He+H\mu $.
As it was already mentioned this
potential has an attractive part and supports the resonant state we are
interested. $Q_{12}(\rho )$ and $P_{12}(\rho )$ give rise coupling of
channels.  In case of small coupling energy levels of $^{4}\!  H\! ed\mu $
and $^{3}\!  H\! ed\mu $ will be found as eigenvalues of the equation:
\begin{equation}
[\frac{d^2}{d\rho
^2}-\frac{15}{4\rho ^2}- \varepsilon _{2}(\rho)-P_{22}(\rho
)+E]u_{2}(\rho )=0
\label{eq:s15}
\end{equation}
for zero boundary conditions
\begin{equation}
u_{2}(0)=u_{2}(\infty )=0.
\label{eq:s16}
\end{equation}
Analogously the continuum wave-function has been found in the
one-level approximation as a solution of
the equation:
\begin{equation}
[\frac{d^2}{d\rho
^2}-\frac{15}{4\rho ^2}- \varepsilon _{1}(\rho)-P_{11}(\rho
)+E]u_{1k}(\rho )=0
\label{eq:s17}
\end{equation}
for the following boundary and asymptotic conditions:
\begin{equation}\left.
\begin{array}{l}
u_{1k}(0)=0,\\
{\displaystyle u_{1k}(\rho )\longrightarrow sin(k\rho+\delta),}
\atop{\hspace{-1cm}\rho \to \infty}
\label{eq:s18}
\end{array}\right.
\end{equation}
where $k=(E-\varepsilon _{1}(\infty ))^{1/2}$ and phase $\delta $
is of no interest for our purposes.
The radiationless decay rate is given by
\begin{equation}
\lambda =\frac{1}{k}\left|M_{k}\right|^{2}\cdot
\displaystyle\frac{me^4}{\hbar^3}s^{-1},
\label{eq:s19}
\end{equation}
where the matrix element of the channel coupling operator is
\begin{equation}
M_{k}=\int\limits_{0}^{\infty }d\rho u_{1k}(\rho )[Q_{12}(\rho )\frac{d}{d\rho}
+
\frac{d}{d\rho} Q_{12}(\rho )-P_{12}(\rho )]u_{2}(\rho ).
\label{eq:s20}
\end{equation}

\section{Numerical results}
The following values of the masses were used in calculations:
$m_\mu=206.769m_e$, $m_d=3670.481m_e$, $m_{^{4}\! H\! e}=7294.295m_e$,
$m_{^{3}\! H\! e}=5495.881m_e$. Equation (\ref{eq:e7}) has been
solved for a number of $\rho $ values in the interval $0\leq \rho \leq 45$.
Variations of the upper bound of this interval do not change final
results. Expressions (\ref{eq:e12})-(\ref{eq:p1}) have been used to
calculate $\varepsilon _{n}(\rho ), Q_{ni}(\rho ), P_{ni}(\rho )$
for these $\rho $ values. The set of trial functions (\ref{eq:s13})
has been adjusted in the following way: numbers of channel-type
functions $N_1$ and hyperspherical harmonics $N_2$ were chosen as:
\begin{equation}\left.
\begin{array}{l}
N_1=2,\  N_2=105,\qquad \rho \leq 5;\\
N_1=6,\  N_2=91,\,\ \qquad  5<\rho <7;\\
N_1=6,\  N_2=105,\qquad  7<\rho \leq 15;\\
N_1=12,\  N_2=78,\qquad  15<\rho .
\label{eq:n21}
\end{array}\right.
\end{equation}
The relative accuracy of two lowest eigenpotentials
$\varepsilon _{1}(\rho ), \varepsilon _{2}(\rho )$ calculated in the
above mentioned interval of $\rho $ can be estimated as $10^{-4}$.
Mesomolecular binding energies $E_L$ and radiationless decay rates
$\lambda _L$ for angular momentum values $L=0,1,2$ have been calculated as
described in the previous section.

The integrand in (\ref{eq:s20}) contains the rapidly
oscillating function $u_{1k}(\rho )$, the sharp functions $u_{2}(\rho )$,
$Q_{12}(\rho ), P_{12}(\rho )$ and their derivatives. In consequence of
these facts, special care has been taken of the evaluation of this
integral.  For this purpose, $u_2(\rho ), Q_{12}(\rho )$ and $P_{12}(\rho )$
were expressed as a product of sharp functions given in the analytical
form and of smooth functions given numerically. A few
per cent variation of decay rates was found when using different ways for
analytical representation of sharp functions.

The calculated values of the binding energies $E_{BL}=E_{d\mu }-E_L$ and
 decay rates for the $^{3,4}\! H\! ed\mu $ systems are presented in Table
1 in comparison with the results of other authors.

{\large {Table 1}
\begin{flushleft}
\begin{tabular}{|c|c|c|c|c|c|c|} \hline system & &
{}~\cite{kino} & ~\cite{hara} & ~\cite{ger} & ~\cite{krav2} & present\\ \hline
  $^{4}\! H\! ed\mu $ & $E_{B0}$ &  & & 77.96 & 78.7 & 77.49 \\ \cline
  {2-7} & $E_{B1}$ & 58.22 & 57.84 & 56.10 & 57.6 & 55.74 \\ \cline {2-7}
  & $E_{B2}$ &  & &  & 20.3 & 17.47\\ \cline {2-7}
  & $\lambda _0$ & & & 2.3 & 1.85 & 0.73 \\ \cline {2-7}
  & $\lambda _1$ & 1.67 & &  2.4 & 1.38 & 1.20 \\ \cline {2-7}
  & $\lambda _2$ & & & & 0.9 & 1.04\\ \hline
  $^{3}\! H\! ed\mu $ & $E_{B0}$ &  &  70.74 & 69.96 & 70.6 & 69.37 \\
  \cline {2-7} & $E_{B1}$ & 48.42 & 47.90 & 46.75 & 48.2 & 46.31 \\ \cline
  {2-7} & $E_{B2}$ &  & &  & 9.6 & 7.11\\ \cline {2-7} & $\lambda _0$ & &
  & 8.0 & 3.58 & 2.87 \\ \cline {2-7} & $\lambda _1$ & 5.06 & & 7.0 & 2.77
  & 3.22 \\ \cline {2-7} & $\lambda _2$ & & & & 1.54 & 1.74\\ \hline
\end{tabular}
\end{flushleft} }
\noindent
{Table 1. Binding energies $E_{BL}~(eV)$ and decay rates
$\lambda_L~(10^{11}s^{-1})$ of the systems $^{3,4}\! H\! ed\mu $
calculated in Ref.~\cite{kino},~\cite{ger},~\cite{krav2},~\cite{hara}
and in the present paper.}

\section{Discussion}

{}From Table 1 it is clear that binding energies for a given $L$ are close
to each other in all calculations. One can see that energies of the
present paper are higher in comparison with
calculations ~\cite{kino} and~\cite{hara}.
The method of this
work gives an upper bound of eigenenergy if the coupling of channels is
omitted. One can conclude that this fact supports the validity of the
one-level approximation in our approach.

The comparison with the
results obtained in the framework of the Born-Oppenheimer approximation
(~\cite{ger},~\cite{krav2}) cannot be done straightforwardly due to the
following reasons. First, mass values and thresholds are introduced
in these calculations {\it ad hoc} and do not coincide with the physical
ones. The importance of these procedures for the
calculation of the decay rate is not clear. Unlike the eigenenergy
problem, the calculation of the decay rate is very sensitive to the fine
details of wave-functions, as is clear from expression (\ref{eq:s20}).
The quasiclassical approximation used in the calculation of the decay rate
in the paper (~\cite{krav2}) can be an origin of an
additional uncertainty.

Qualitatively, all calculations support the strong isotopic dependence of
the decay rates observed in experiment (\cite{mats},~\cite{ishid})
Nevertheless, the calculated values are quite different and consistency
of theoretical results should be reached.

It is accepted that the formation of $H\! ed\mu $ molecules takes place
in the state with $L=1$. In this connection,
for comparison with experiment, the most important is the ratio
$\lambda_\gamma/(\lambda_\gamma+\lambda_1)$, where $\lambda_\gamma $ is
the radiative decay rate from the molecular state $L=1$. Using
$\lambda_\gamma $ from the paper~\cite{hara} and the present values of
$\lambda_1$, one comes to the ratio
$\lambda_\gamma/(\lambda_\gamma+\lambda_1)=0.585$ for $^{4}\! H\! ed\mu
$ and $\lambda_\gamma/(\lambda_\gamma+\lambda_1)=0.325$ for the $^{3}\! H\!
ed\mu $ systems.
Other processes, which may be important in the experiment, are collisional
transitions to the muonic molecular states with angular momentum $L\neq 1$.

Finally, one would like to emphasize the necessity of the systematic study
in the framework of the same approach of the processes involved in the
formation, rearrangement and decay of systems under consideration.

\section{Acknowledgment}
One of the authors, V.B.Belyaev, would like to thank  the  Scientific
Division of NATO for the financial support  within  the  Collaborative
Research Grant No.~930102.


\end{document}